\documentclass[prd,showpacs]{revtex4}

\usepackage{amsmath}
\usepackage{amsthm}
\usepackage{amssymb}
\usepackage{amscd}
\usepackage[british]{babel}
\usepackage{graphicx}
\usepackage{psfrag}
\usepackage{epsfig}
\usepackage{rotating}
\usepackage{times}

\begin{document}

\title{Sudden singularities in semiclassical gravity}

\author{Jaime de Haro$^{1,}$\footnote{E-mail: jaime.haro@upc.edu},
Jaume Amoros$^{1,}$\footnote{E-mail: jaume.amoros@upc.edu} and
Emilio Elizalde$^{2,}$\footnote{E-mail: elizalde@ieec.uab.es,
elizalde@math.mit.edu}}

\affiliation{$^1$Departament de Matem\`atica Aplicada I, Universitat
Polit\`ecnica de Catalunya, Diagonal 647, 08028 Barcelona, Spain \\
$^2$Instituto de Ciencias del Espacio (CSIC) \& Institut
d'Estudis Espacials de Catalunya (IEEC/CSIC)\\ Campus UAB, Facultat
de Ci\`encies, Torre C5-Parell-2a planta, 08193 Bellaterra
(Barcelona) Spain}

\pagestyle{myheadings}

\theoremstyle{plain}
\newtheorem{lemma}{Lemma}[section]
\newtheorem{theorem}{Theorem}[section]
\newtheorem{proposition}{Proposition}[section]
\newtheorem{corollary}{Corollary}[section]
\newtheorem{remark}{Remark}[section]
\newtheorem{definition}{Definition}[section]
\newtheorem{example}{Example}[section]

\newcommand{\boxend}{\flushright{$\Box$}}
\newenvironment{dem}[1]{\begin{trivlist} \item {\bf Proof #1\/:}}
            {\boxend \end{trivlist}}

\newcommand{\N}{{\mathbb N}}               
\newcommand{\Z}{{\mathbb Z}}               
\newcommand{\Q}{{\mathbb Q}}               
\newcommand{\R}{{\mathbb R}}               
\newcommand{\C}{{\mathbb C}}               
\renewcommand{\S}{{\mathbb S}}             
\newcommand{\T}{{\mathbb T}}               
\newcommand{\D}{{\mathbb D}}               

\newcommand{\half}{\frac{1}{2}}
\renewcommand{\Re}{\mbox{\rm Re}}
\renewcommand{\Im}{\mbox{\rm Im}}
\newcommand{\sprod}[2]{\left\langle#1,#2\right\rangle}
\newcommand{\inner}[2]{\left\langle#1,#2\right\rangle_2}

\newcommand{\e}{\epsilon}
\newcommand{\w}{\omega}
\newcommand{\f}{\frac}

\newcommand{\ep}{\varepsilon}
\newcommand{\al}{\alpha}
\newcommand{\h}{\hbar}
\renewcommand{\tilde}{\widetilde}

\thispagestyle{empty}

\begin{abstract}
It has been claimed in a recent paper
\cite{bbfhd12} that sudden singularities will survive in semiclassical gravity. This issue is here carefully reviewed, pointing out that
such conclusion, even if valid under some specific conditions, does not stand in other cases. An explicit example is studied in detail to support our statement, reached in these other situations,
that quantum effects may in fact drastically modify the behavior of sudden singularities.
\end{abstract}


\maketitle

\vspace{-6.5mm}

\hspace*{10mm}  {\footnotesize Keywords: Future
sudden singularities, semiclassical gravity}

\section{Introduction}
Since the discovery that our universe is expanding in an accelerated way, a number of papers
discussing  models with phantom dark energy have been proposed in order to explain this unexpected behavior
 (see, for instance, \cite{ckw03,not05}). These models have the common feature that future singularities show up,
like (big rip and sudden ones, and others) \cite{ckw03,b04}. Those could, in some cases, be avoided using semiclassical gravity.
Indeed, it has been shown explicitly in a number of papers that these singularities are modified when one takes
into account quantum effects due to massless, conformally coupled fields \cite{no04,not05,no04a, noe04, hae11,ha11}.
In a recent work \cite{bbfhd12} (see also the preprint \cite{bbfhd12a}) it has been claimed however that, quite on the contrary, quantum effects produced by a massive conformally coupled field cannot prevent the formation of sudden singularities, the
reason being that the renormalized energy of the created particles is zero and, thus, that  backreaction of the quantum
effects will not change  the evolution of the Universe near the future singularity. In the present paper we review in some
detail and try to understand the results obtained in \cite{bbfhd12}. It will be argued that
renormalization of the energy density can in fact be performed in a different way, which seems reasonable enough and leads as a result to an energy density different from zero.
Consequently, it can be concluded that there may be a backreaction available to modify the classical evolution of our Universe. In support of our conclusion we will here present a
specific example with sudden singularities, where a sudden singularity is drastically modified when one takes into account quantum effects due to the conformally coupled fields, in the renormalization scheme adopted through this paper.

\section{Summary and discussion of previous results}
In this Section we will review, using a standard language, the work done in Ref.~\cite{bbfhd12}. The units employed throughout the paper are $c=\hbar=M_p=1$, being $M_p$ the reduced Planck mass. In \cite{bbfhd12} a model was considered where the sudden singularity is described by
the following scalar factor \cite{b04}:
\begin{equation}\label{A1}
a(t)=\left(\frac{t}{t_s}\right)^{1/2}(a_s-1)+1-\left(1-\frac{t}{t_s}\right)^n,
\end{equation}
where $t_s$ is the time at which the sudden singularity occurs, $a_s$ is the value of the scalar factor at that time, and $n$ is a parameter, satisfying $1<n<2$.
When  $t\rightarrow 0$ the universe is in the radiation phase $\left[ a\rightarrow \left(\frac{t}{t_s}\right)^{1/2}(a_s-1)\right]$, and when
$t\rightarrow t_s$ the universe is in the singular phase $\left[a\rightarrow a_s,\quad
\dot{a}\rightarrow \frac{1}{2t_s}(a_s-1),
 \mbox{ and }\ddot{a}\rightarrow -\frac{n(n-1)}{t_s^2}
\left(1-\frac{t}{t_s}\right)^{n-2}\right]$.

On the other hand, the modes corresponding to a conformally coupled, massive scalar field, $\phi$, satisfy the Klein-Gordon equation
\begin{equation}\label{A2}
 \chi_k''+[k^2+m^2 a^2(\eta)]\chi_k=0,
\end{equation}
 where $\eta$ is the conformal time, $\chi=\phi a$, and $\chi'$ its derivative with respect to the conformal time. One can see that, in the radiation phase ($\eta\rightarrow 0$), Eq.~(\ref{A2}) becomes  $\chi_k''+k^2\chi_k=0,$ and that, in the singular phase ($t\rightarrow t_s$), it reduces to $\chi_k''+(k^2+m^2 a_s^2)\chi_k=0.$

Further, since  equation (\ref{A2}) is not solvable analytically, in Ref.~\cite{bbfhd12} it is approximated by the more simple one
\begin{equation}\label{A3}
 \chi_k''+\omega_k^2(\eta)\chi_k=0,
\end{equation}
with
\begin{equation}\label{A4}
 \omega_k^2(\eta)=\left\{\begin{array}{ccc}
 k^2,& \mbox{for} & \eta<\eta_c, \\
  k^2+m^2a_s^2,& \mbox{for} & \eta>\eta_c,\end{array}
\right.
\end{equation}
$\eta_c$ being the time at which the sudden transition occurs.
Note that, this choice for the frequency is equivalent to selecting as scale factor in Eq.~\ref{A2})
\begin{equation}
 a(\eta)=\left\{\begin{array}{ccc}
 0,& \mbox{for} & \eta<\eta_c, \\
  a_s,& \mbox{for} & \eta>\eta_c,\end{array}
\right.
\end{equation}
which is discontinuous at $\eta=\eta_c$.

Eq.~(\ref{A3}), with the frequency (\ref{A4}), has the solution
\begin{equation}\label{A5}
 \chi_k(\eta)=\left\{\begin{array}{ccc}
 e^{-ik\eta}/\sqrt{2k},& \mbox{for} & \eta<\eta_c, \\ & & \\
  \alpha_ke^{-i\tilde{\omega}_k\eta}/\sqrt{2\tilde{\omega}_k}+
  \beta_ke^{i\tilde{\omega}_k\eta}/\sqrt{2\tilde{\omega}_k},
 & \mbox{for} & \eta>\eta_c,\end{array}
\right.
\end{equation}
being $\alpha_k$ and $\beta_k$ Bogoliubov coefficients, and $\tilde{\omega}_k=\sqrt{ k^2+m^2a_s^2}$.
Matching at $\eta_c$, one obtains
\begin{equation}\label{A6}
 \beta_k=\frac{m^2a_s^2}{2\sqrt{k\tilde{\omega}_k}(k+\tilde{\omega}_k)},
\end{equation}
thus, the energy of each mode is
\begin{equation}\label{A7}
 \rho_{part,k}(m)=\tilde{\omega}_k|\beta_k|^2=\frac{m^4a_s^4}{4{k}(k+\tilde{\omega}_k)^2}.
\end{equation}
It follows from this result that the energy density of the particles produced in the process [see Eq.~(5.112) of \cite{bd82}], namely
\begin{equation}\label{A8}
\rho_{par}(m)=\frac{1}{2\pi^2 a_s^4}\int_0^{\infty} dk k^2 \rho_{part,k}(m),
\end{equation}
diverges. What is here actually important to stress is the fact that the energy density could diverge
due to the sudden transition. Had the transition been not so abrupt---what indeed occurs with the 
original equation (\ref{A2})---the density of produced 
particles could have remained finite (a clear example of this phenomenon can be found in \cite{f87}, see also \cite{h10}) .

In \cite{bbfhd12} the renormalization of Eq.~(\ref{A8}) is undertaken. To do that the authors use
 the so-called $n$-wave procedure \cite{zs72,gmm94,blm98}, apparently in an incomplete way.
To wit, this method is used to renormalize the energy density tensor whose first component is
the total energy density. For the case of a conformally coupled,  massive scalar field this is (see, for instance, Eq.~(9.152) of \cite{gmm94}):
\begin{equation}\label{A9}
\rho(m)=\rho_{par}(m)+\rho_{vac}(m)=\frac{1}{2\pi^2 a_s^4}\int_0^{\infty} dk k^2 \rho_{part,k}(m)+ \frac{1}{2\pi^2 a_s^4}\int_0^{\infty} dk k^2
\frac{\tilde{\omega}_k}{2},
\end{equation}
and corresponds to the energy density of the particles produced in the process plus the energy density of the zero-point oscillations of the vacuum.

In fact, this procedure is easy to explain: defining $\rho^{(n)}_k(m)\equiv \frac{1}{n}\rho_{nk}(nm)$ where $\rho_{k}(m)=\rho_{part,k}(m)+\frac{\tilde{\omega}_k}{2}$, the renormalized energy per mode is given by \cite{zs72}
\begin{equation}\label{A10}
\rho_{k}^{ren}(m)=\lim_{n\rightarrow 0}\left[ \rho_{k}(m)-\rho^{(n)}_k(m)-\frac{\partial}{\partial(n^{-2})}\rho^{(n)}_k(m)-
\frac{1}{2}\frac{\partial^2}{\partial(n^{-2})^2}\rho^{(n)}_k(m)
\right],
\end{equation}
and,  applying the method to $\rho_{k}(m)=\rho_{part,k}(m)+\frac{\tilde{\omega}_k}{2}$, 
one obtains $\rho_{k}^{ren}(m)=0$, because $\rho^{(n)}_k(m)=\rho_{k}(m)$.

However, there is another way to renormalize the energy density, which we will follow here. 
Since this method was  introduced, at the beginning, for smooth transitions, that is, when the frequency $\omega_k=\sqrt{k^2+m^2a^2(\eta)}$
is a {\it smooth} function--what does not happen for the frequency defined in (\ref{A4})--when the transition is abrupt we will not apply directly expression (\ref{A10}). In that case, we go back to the general formulae (see, e.g., Eqs.~(9.161)-(9.164) of \cite{gmm94}
or Eqs.~(29)-(31) of \cite{blm98}), which have a wider range of applicability. In the case at hand, one just needs to subtract the energy of the zero-point oscillations of
the vacuum $\frac{\tilde{\omega}_k}{2}$, which correspond in the general situation to the term $\rho^{(n)}_k(m)$ (see, for instance,  the paragraph below formula (21) of \cite{zs72}), thus obtaining
\begin{equation}\label{A11}
\rho_{k}^{ren}(m)=\rho_{part,k}(m),
\end{equation}
which, indeed, corresponds to the energy of the produced particles in the $k$-mode. As a consequence,
the energy density of the particles produced remains divergent. But this could be due to the 
assumption that the transition from the radiation phase to the singular one is abrupt. Anyway, what is clear is that,
 as a result of our procedure, the renormalized energy density is no more zero.

As an interesting remark we note that one can alternatively obtain the same result (\ref{A11}) by employing the adiabatic substraction
prescription  \cite{pf73}. In fact, it is strightforward to check this extreme using Eqs.~(3.1), (3.7), (3.11), and (3.14) of \cite{b80}.

Further to the point, another difference appear with our procedure, stemming from the fact that with the method used in  \cite{bbfhd12}
there is no vacuum polarization effect. To be specific, let us consider the scalar, 
massless, conformally coupled case, where the well-known renormalized energy density is given by (see, e.g., \cite{d77, ha11})
\begin{eqnarray}\label{A12}
\rho^{ren}(0)=\frac{1}{480\pi^2}(3H^2\dot{H}+H\ddot{H}-\frac{1}{2}\dot{H}^2)+\frac{1}{960\pi^2}
H^4.
\end{eqnarray}
Taking further into account that $p^{ren}(0)=\frac{1}{3}(\rho^{ren}(0)-T^{ren}(0))$, with
\begin{eqnarray}\label{1b}T^{ren}(0)=\frac{1}{480\pi^2}
(\dddot{H}+12H^2\dot{H}+7H\ddot{H}+4\dot{H}^2)+\frac{1}{240\pi^2}
(H^4+H^2\dot{H})\end{eqnarray}
 the anomalous trace \cite{ha11}, one obtains de renormalized pressure
\begin{eqnarray}\label{A13}
p^{ren}(0)=-\frac{1}{1440\pi^2}(\dddot{H}+11H^2\dot{H}+6H\ddot{H}+\frac{9}{2}\dot{H}^2)-\frac{1}{960\pi^2}
H^4.
\end{eqnarray}
If one takes the  scalar factor to be constant at late times, $a(t)=a_s$, then $\rho^{ren}(0)$ and  $p^{ren}(0)$ will vanish. But, introducing
 the scalar factor as given by (\ref{A1}), one gets
\begin{eqnarray}\label{A14}
\rho^{ren}(0)\sim \frac{1}{960\pi^2}\frac{n(n-1)(n-2)(a_s-1)}{a^2_st^4_s}\left(1-\frac{t}{t_s}\right)^{n-3},
\end{eqnarray}
and
\begin{eqnarray}\label{A15}
p^{ren}(0)\sim \frac{1}{1440\pi^2}\frac{n(n-1)(n-2)(n-3)}{a^4_st^4_s}\left(1-\frac{t}{t_s}\right)^{n-4},
\end{eqnarray}
which are divergent quantities at $t=t_s$. Consequently, they may drastically
change the behavior of the scale factor at $t=t_s$. In other words, quantum effects might actually modify the singularity.
More precisely, consider the Friedmann, $H^2=\frac{1}{2}\rho$, and the Raychaudury equation, $\dot{H}=-\frac{1}{2}(\rho+p)$.
Near the singularity, one has
\begin{eqnarray}\label{A16}
\rho\sim \frac{3(a_s-1)}{a^2_st^2_s}\left[\frac{(a_s-1)}{4}+n\left(1-\frac{t}{t_s}\right)^{n-1}\right], \qquad
p\sim \frac{2n(n-1)}{a_st^2_s}\left(1-\frac{t}{t_s}\right)^{n-2}.
\end{eqnarray}
From here, proceeding as in \cite{no04}, we consider the Friedmann semiclassical equation
\begin{eqnarray}\label{A17}
 H^2=\frac{1}{3}[\rho+\rho^{ren}(0)],
\end{eqnarray}
where $\rho$ is given by (\ref{A16}), and then look for singular solutions of this semiclassical equation, with the form
\begin{equation}\label{A18}
H(t)=H_s-C\left(1-\frac{t}{t_s}\right)^{n'},
\end{equation}
where $H_s$, $C$ and $n'$ are unknown parameters. Inserting (\ref{A18}) into  (\ref{A17}) and retaining the leading terms, we obtain
\begin{eqnarray}\label{A19}
 n'=n+1, \qquad H_s= 12\sqrt{10}\pi\sqrt{1+\sqrt{1-\frac{(a_s-1)^2}{8640\pi^2a_s^2t_s^2}}}, \qquad
C=\frac{480\pi^2 (a_s-1)}{H_sa_s(n+1)}.
\end{eqnarray}
Then, since $3<n'<4$, it turns out that $\dot{H}$, and $\ddot{H}$ do not diverge at $t=t_s$, which means that, for this kind of singular solutions, the singularity becomes much milder, owing to the quantum corrections. In fact, if one does not take into account these quantum corrections, i.e., using Eq.~(\ref{A1}), one easily sees that $\dot{H}$ diverges.

To show even more clearly that quantum corrections may drastically alter, in some cases, the behavior of future singularities we will consider,
in the next Section, a specific example that can be studied
qualitatively using the theory of dynamical system.

\section{A model driving sudden singularities}
 In this Section we study in detail the solutions of the model, inspired in equations $(21)$ and $(59)$ of \cite{not05},  given by the equation of state
 $p=-\rho-\frac{\rho\rho_i^{\gamma}}{(\rho_s-\rho)^{\gamma}}$, with $\gamma>0$ and
 $\rho_s, \rho_i$  positive parameters, which also gives rise to a sudden singularity, because the pressure diverges at $\rho=\rho_s$ (a finite value of the energy density).
 To simplify the calculations, we will set $\gamma=3$ and $\rho_s= \rho_i$. Then,
solving the classical Friedmann and continuity equations
\begin{eqnarray}\label{b2}
 H^2=\frac{\rho}{3},\quad \dot{\rho}=-3H(p+\rho),
\end{eqnarray}
one obtains
\begin{eqnarray}
 \rho\sim \rho_s\left[1-\left(48{\rho_s}\right)^{1/8}(t_s-t)^{1/4}\right],\quad &&
H\sim \sqrt{\rho_s/3}\left[1-\frac{1}{2}\left(48{\rho_s}\right)^{1/8}(t_s-t)^{1/4}\right]\nonumber \\&&
p\sim -{\rho_s}\left(48{\rho_s}\right)^{-3/8}(t_s-t)^{-3/4},
\end{eqnarray}
which proves that, in this model, the universe develops a sudden singularity at time $t=t_s$.

Note however that this model is very different to the one discussed in Section II. Effectively, the model proposed in \cite{bbfhd12,b04} satisfies the
strong energy-condition $\rho>0$ and $\rho+3p>0$. On the other hand, in this Section we propose a universe containing phantom dark energy
($\rho+p=-\frac{\rho\rho_i^{\gamma}}{(\rho_s-\rho)^{\gamma}}<0 $). 

Considering once again a scalar massless, conformally coupled field, the anomalous trace is given by \cite{hae11}
\begin{equation}\label{C1}
 T^{ren}(0)=\frac{1}{2880\pi^2}\square R+\frac{1}{5760\pi^2}G,
\end{equation}
where $R=6(\dot{H}+2H^2)$ is the scalar curvature and $G=24H^2(H^2+\dot{H})$ the Gauss-Bonnet curvature. In terms of the Hubble parameter, Eq.~(\ref{C1}) becomes Eq.~(\ref{1b}). To obtain the energy density, one has to introduce the trace anomaly $T^{ren}(0)=\rho^{ren}(0)-3p^{ren}(0)$
into the conservation equation $\dot{\rho}^{ren}(0)=-3H(\rho^{ren}(0)+p^{ren}(0))$ to get
$\dot{\rho}^{ren}(0)=-H(4\rho^{ren}(0)-T^{ren}(0))$. Since the value of $T^{ren}(0)$ is given by Eq.~(\ref{1b}), this last form of the conservation
condition is a linear first order differential equation that can be easily integrated and whose solution is given by Eq.~(\ref{A12}).

Having obtained the renormalized energy density, the semiclassical Friedmann equation is given by Eq.~(\ref{A17}), which in terms of the Hubble parameter has the form
\begin{equation}\label{C2}
 H^2=\frac{\rho}{3}+\frac{1}{1440\pi^2}(3H^2\dot{H}+H\ddot{H}-\frac{1}{2}\dot{H}^2)+\frac{1}{2880\pi^2}
H^4.
\end{equation}
This is a second order differential equation on $H$ and its solutions could differ from the ones obtained form the classical Friedmann equation.
To show this we write the semiclassical Friedmann equation and the conservation equation as an autonomous system, which general solution is a two-parameter family
\begin{eqnarray}\label{A23}\left\{\begin{array}{ccl}
{\bar{H}}'&=&\bar{Y},\\
{\bar{Y}}'&=& \frac{1}{2\bar{H}}\left({\bar{H}^2}-\bar{\rho}
-6\bar{H}^2\bar{Y}+\bar{Y}^2-\bar{H}^4\right),\\
{\bar{\rho}}'&=&3\bar{H}\frac{\bar{\rho}\bar{\rho}_s^3}{(\bar{\rho}_s-\bar{\rho})^3},
\end{array}\right.\end{eqnarray}
or, equivalently,
\begin{eqnarray}\label{A24}\left\{\begin{array}{ccl}
\frac{d\bar{Y}}{d\bar{H}}&=& \frac{1}{2\bar{H}\bar{Y}}\left({\bar{H}^2}-\bar{\rho}
-6\bar{H}^2\bar{Y}+ \bar{Y}^2-\bar{H}^4\right),\\
\frac{d\bar{\rho}}{d\bar{H}}&=&3\frac{\bar{H}}{\bar{Y}}
\frac{\bar{\rho}\bar{\rho}_s^3}{(\bar{\rho}_s-\bar{\rho})^3} ,\end{array}\right.\end{eqnarray}
where we have introduced the dimensionless variables $\bar{t}=H_+t$, $\bar{H}=H/H_+$,
$\bar{Y}=Y/H_+^2$, and $\bar{\rho}=\frac{\rho}{3H_+^2}$, being $H_{+}=\sqrt{2880\pi^2}$, while $'$ denotes derivative with respect to the time $\bar{t}$.

In the same way as it was proven in \cite{hae11}, in the contracting region ($\bar{H}<0$) the system (\ref{A23}) has,
 for general solution, a two-parameter family of future-singular solutions, namely
\begin{eqnarray}\label{A25}
 \bar{H}\sim \frac{-3(1\pm \sqrt{2/3})}{\bar{t}_s-\bar{t}}, \quad \bar{\rho}\sim 0.
\end{eqnarray}
In the expanding phase ($\bar{H}>0$), we look for future singular behaviors of the form
$\bar{\rho}=\bar{\rho}_s-\bar{\rho}_1(\bar{t}_s-\bar{t})^{\mu}$
with $\mu>0$, $\bar{\rho}_1>0$ and $\bar{t}_s>\bar{t}$. Near
$\bar{t}_s$, from (\ref{A23}) and retaining the leading terms only, we get the particular solution
\begin{eqnarray}
\bar{\rho}\sim\bar{\rho}_s\left[1-(36\bar{\rho}_s)^{1/8}(\bar{t}_s-\bar{t})^{1/2}\right]
\quad\mbox{and}\quad
\bar{H}\sim \sqrt{\bar{\rho}_s}(\bar{t}_s-\bar{t})\left[1-\frac{2}{3}(36\bar{\rho}_s)^{1/8}
(\bar{t}_s-\bar{t})^{1/2}\right],
\end{eqnarray}
what means that both not only $\bar{H}$,  but also $\bar{H}'$, are finite near
$\bar{\rho}_s$, after the quantum corrections are duely taken into account.

This means that, when one incorporates quantum corrections the singularity becomes softer in the expanding phase ($\bar{H}>0$). Note also that, at the singularity time we have $\bar{H}(\bar{t}_s)=0$, consequently, this particular solution does not enter the contracting phase.

On the other hand, the solutions without future singularities are of the same kind as the ones that appear when one considers a universe with a big rip modeled by the equation of state $p=\omega\rho$, with $\omega<-1$, and one takes into account quantum corrections
(see, for instance, \cite{hae11}). These solutions are given by a one-parameter
family in the contracting Friedmann phase plus a particular
solution that converges towards the contracting de Sitter one. To prove this, we consider the Friedmann phase (the classical Friedmann solution) given, in dimensionless variables, by
\begin{eqnarray}\label{c1}
\bar{\rho}=\bar{H}^2,\quad \bar{Y}=\frac{3}{2}\frac{\bar{H}^2\bar{H}_s^6}{(\bar{H}_s-\bar{H})^6},
\end{eqnarray}
where $\bar{H}_s\equiv \sqrt{\bar{\rho}_s}$. Then, for $|\bar{H}|\ll 1$, by linearizing the system
(\ref{A24}) around the classical Friedmann solution (\ref{c1}) and applying  the
WKB method (see, e.g., p.~$53-55$ of \cite{fedo87}) one gets the following one-parameter family of
solutions which converges to the classical Friedmann solution, when  $\bar{H}\rightarrow 0$,
\begin{eqnarray}\left(\begin{array}{c}
\bar{Y}\\\\\bar{\rho}
\end{array}
\right) \cong
\left(\begin{array}{c}\frac{3}{2}\frac{\bar{H}^2\bar{H}_s^6}{(\bar{H}_s-\bar{H})^6}
\\\\\bar{H}^2
\end{array}
\right)+K\frac{e^{\frac{2}{3\bar{H}}}}
{\sqrt{-\bar{H}}}\left(\begin{array}{c}
1\\\\ 2\bar{H}
\end{array}
\right),\quad \mbox{for}\quad \bar{H}<0,
\end{eqnarray} being $K$ a free parameter. To prove that there is a particular solution which tends towards the contracting de Sitter phase, one must
note that $(-1,0,0)$ is a critical point of (\ref{A23}) (in fact, it is the contracting de Sitter solution). Then, linearizing this system around that point, one obtains a matrix with eigenvalues $\lambda_1=-3<0$ and $\lambda_{\pm}=\frac{3\pm\sqrt{5}}{2}>0$, what finally proves the statement.

As last point, in order to qualitatively study the system, it is convenient to perform the change of variable $\bar{p}=\sqrt{|H|}$. After this, the semiclassical Friedmann equation becomes
\begin{eqnarray}\label{C4}
\frac{d}{d\bar{t}}\left((\bar{p}^{'})^2/2+{V}(\bar{p},\bar{\rho})\right)=-3\epsilon
p^2(\bar{p}^{'})^2-\frac{3\epsilon}{8}\frac{\bar{\rho}\bar{\rho}^3_s}{(\bar{\rho}_s-\bar{\rho})^3},
\end{eqnarray}
or, equivalently,
\begin{eqnarray}\label{C3}
\bar{p}^{''}=-\partial_{\bar p}{V}(\bar{p},\bar{\rho})-3\epsilon
p^2\bar{p}^{'},
\end{eqnarray}
where
${V}(\bar{p},\bar{\rho})=-\frac{1}{8}\left(\bar{p}^2(1-\frac{1}{3}\bar{p}^4)+\frac{\bar{\rho}}{
\bar{p}^2}\right)$, and $\epsilon\equiv$ sign$(H)$.

Eq.~(\ref{C4}) shows that the system is dissipative (resp. anti-dissipative) in the expanding phase $H>0$
(resp. in the contracting phase $H<0$) which means that the system loses (resp. gains) energy in the
expanding (resp. contracting) phase. As a consequence of this, when the universe is in the expanding phase it loses energy and,
due to the form of the potential $V$, it rolls down to $\bar{p}=0=\bar{H}$ and enters the contracting phase $\bar{H}<0$, where it
gains enough energy to arrive to $\bar{H}=-\infty$ in a finite time, exhibiting the behavior described by Eq.~(\ref{A25}) (a detailed account of this process is given in \cite{hae11}).

Here one can directly see the main difference between semiclassical and classical cosmology. In semiclassical cosmology, due to the quantum corrections,
  our universe, which nowadays is in the expanding phase, will bounce and will
enter into the contracting one, where it will develop a singular behavior
described by $(28)$. However, in classical cosmology, the expanding and contracting phases decouple, that is, the universe cannot bounce. This is due to the fact that the classical Friedmann equation prescribes selecting $H=\pm\sqrt{\frac{\rho}{3}}$, and then, once the sign has been chosen, it cannot be changed any more.

In order to further check this qualitative analysis, we have performed an accurate numerical study, as shown in Figs.~1 and 2.
\begin{figure}[htb]
\begin{center}
\includegraphics[scale=0.40]{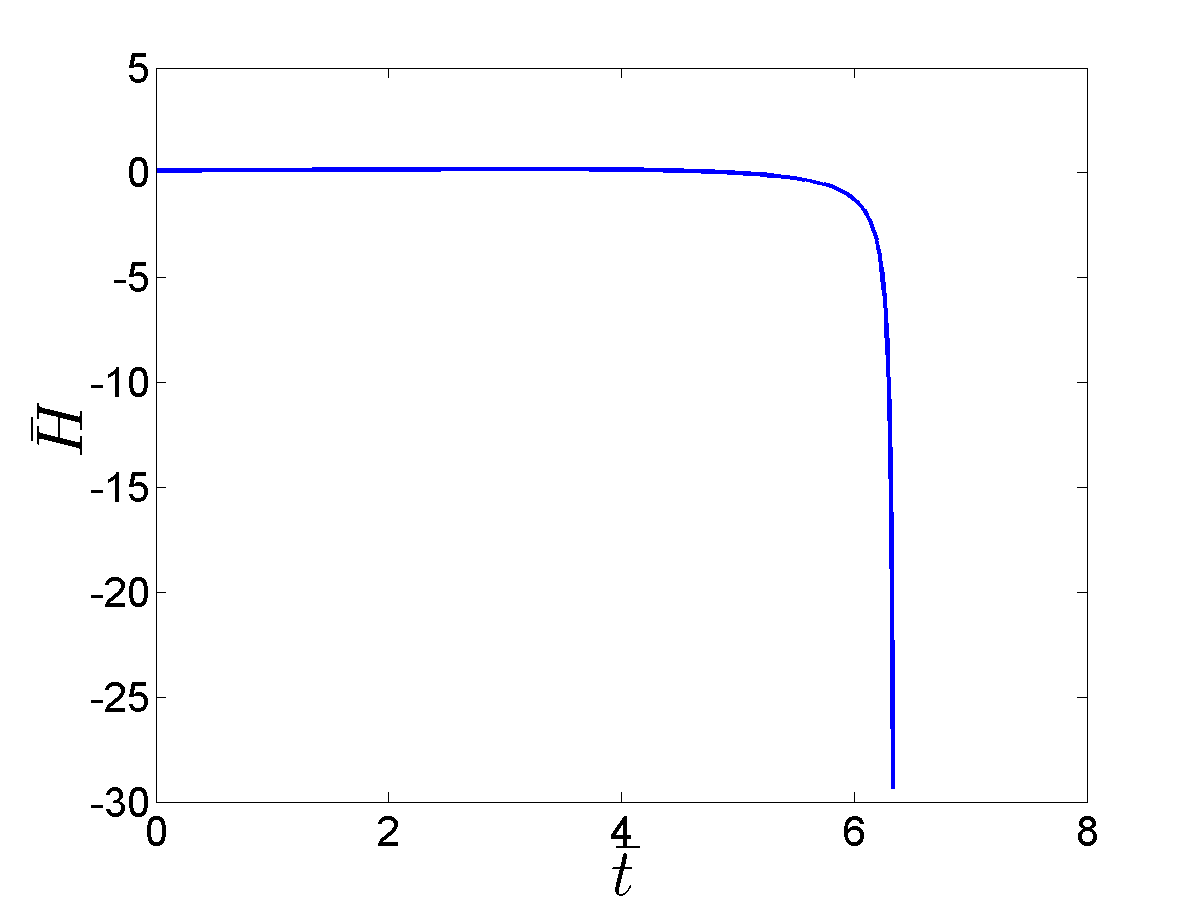}
\includegraphics[scale=0.40]{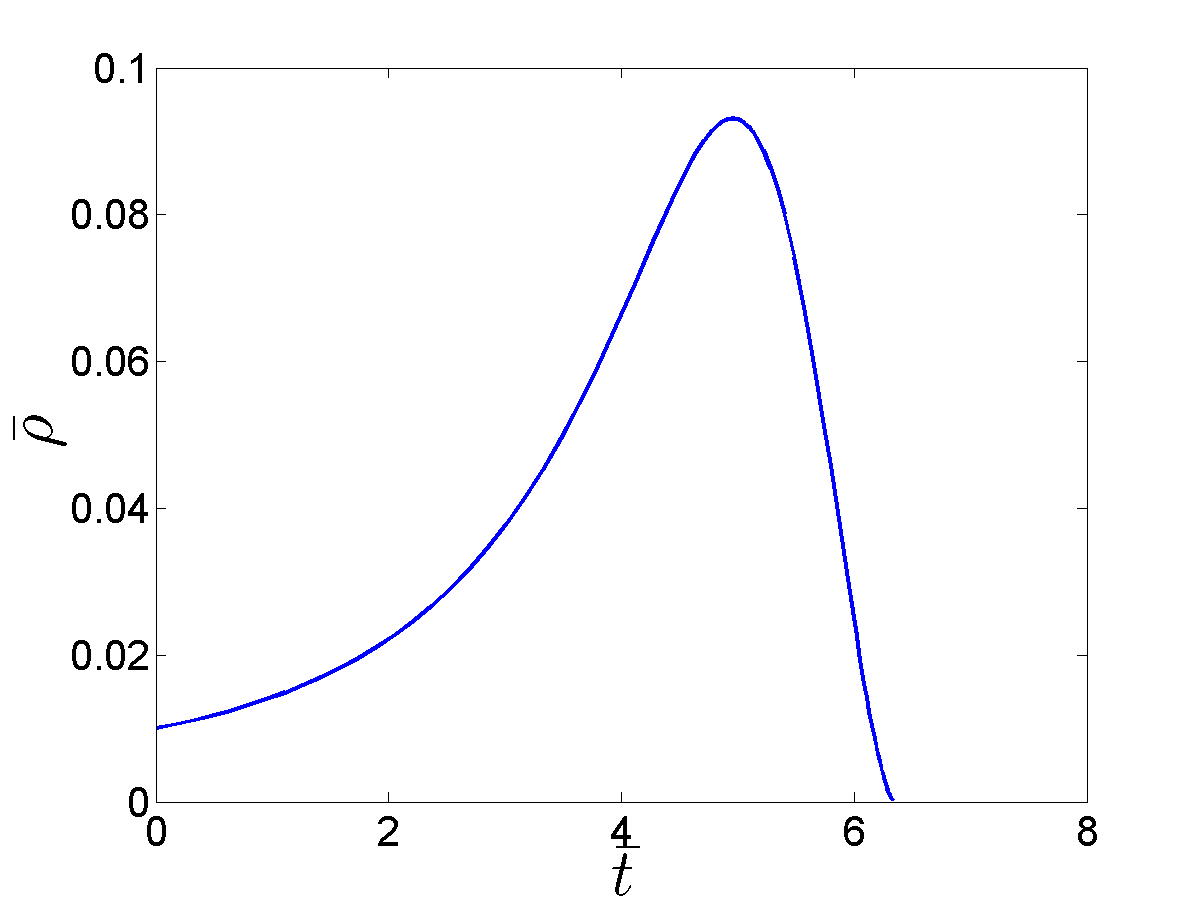}
\end{center}
\caption{$\bar H(\bar t)$ and $\bar \rho(\bar t)$ obtained by integration of (\ref{A23}) with initial
conditions in the Friedmann phase, i.e., with $(\bar{H}_0,\frac{3}{2}\frac{\bar{H}_0^2\bar{H}_s^6}{(\bar{H}_s-\bar{H}_0)^6},\bar{H}_0^2)$, being $\bar{H}_0=0.1$
and $\bar{\rho}_s=1$. Quantum correction drive the universe to the contracting phase, where the universe has the behavior described by Eq.~(\ref{A25}).}
\end{figure}
\begin{figure}[htb]
\begin{center}
\includegraphics[scale=0.60]{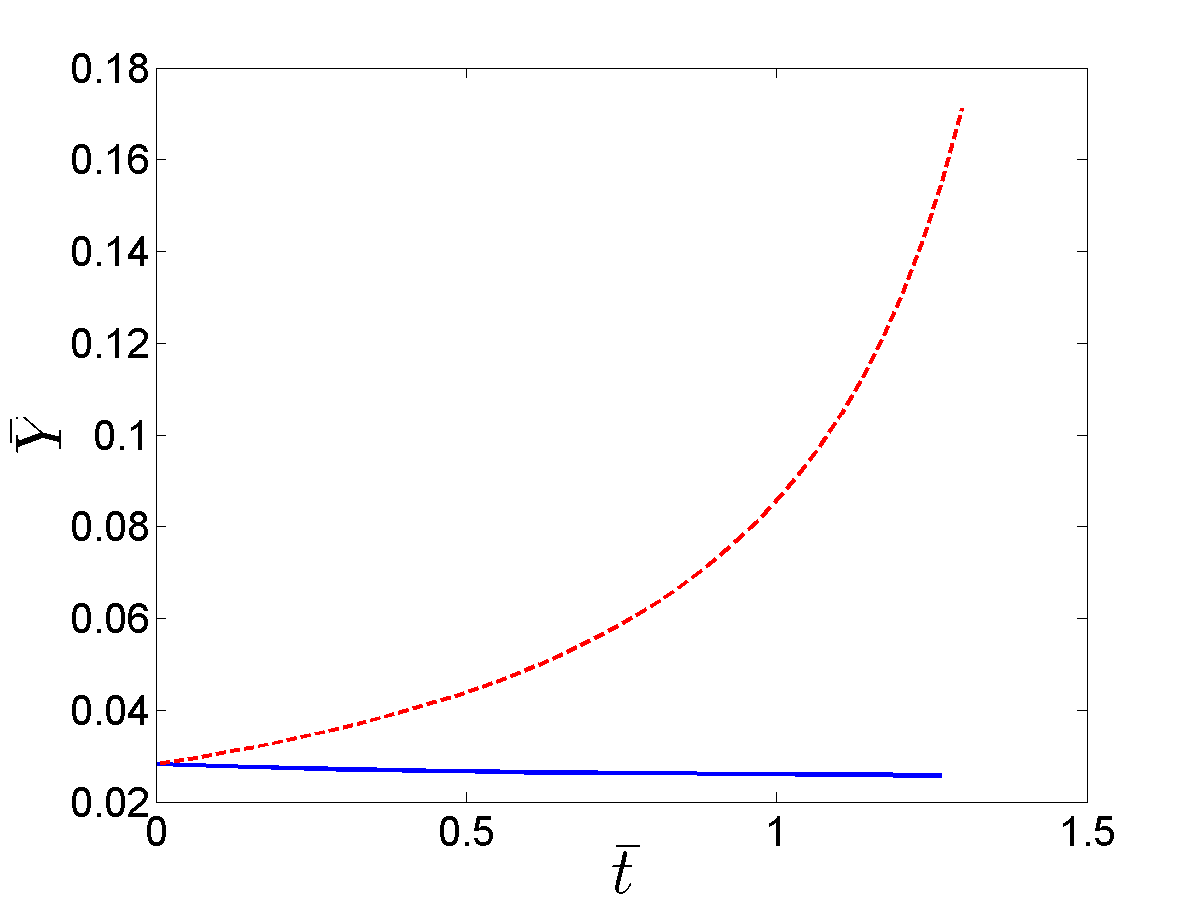}
\end{center}
\caption{Comparison of the derivatives of $\bar H(\bar t)$. The red line corresponds to the derivative of $\bar H(\bar t)$ without quantum corrections (we see that
it will diverge at finite time). The blue line correspond to the derivative of $\bar H(\bar t)$ after taking into account the quantum corrections, as obtained
by numerical integration of the system (\ref{A23}). It is plain from the plot that quantum corrections modify the future singularity.}
\end{figure}

To finish, the above results lead to the conclusion that, while almost any solution will
develop one or more singularities in the contracting phase, the corresponding
scalar factor and  energy density will go down to zero in finite time, as shown in the
specific example of the two-parameter family of solutions given in Eq.~(\ref{A25}).
Note, moreover, that all remaining solutions actually form a zero-measure set of the whole set of solutions. That is, they are unstable, in the sense that any small perturbation of their initial conditions will give rise to a solution which develops future singularities in the contracting phase.
This shows that, in the example considered, quantum effects will drastically modify the expansion of the Universe, along with other physical consequences associated with the presence of classical sudden singularities.


\vspace {5mm}

\noindent{\bf Acknowledgments.} This investigation has been
supported in part by MICINN (Spain), Projects MTM2011-27739-C04-01,
MTM2009-14163-C02-02, FIS2006-02842 and FIS2010-15640, and Contract PR2011-0128,
by CPAN Consolider Ingenio Project, and by AGAUR (Generalitat de Ca\-ta\-lu\-nya),
Contracts 2009SGR 345, 994 and 1284. EE's research was partly carried out while
 on leave at the Department of Physics and Astronomy,
Dartmouth College, 6127 Wilder Laboratory, Hanover, NH 03755, USA.

\end{document}